\documentclass[a4paper,12pt]{article}
\pdfoutput=1
\usepackage{graphicx, rotating,amssymb,amsmath}
\usepackage{cite}

\ifx\pdfoutput\undefined
\usepackage[dvips,bookmarks]{hyperref}	% This is for arXiv.org
\else
\usepackage{hyperref}	% This is for pdftex
\fi
\hypersetup{colorlinks,bookmarksopen,bookmarksnumbered,citecolor=verdeCP3,
linkcolor=bluCP3,pdfstartview=FitH,urlcolor=rossoCP3}
\def\myurl#1#2{\href{http://#1}{#2}}
\def\hhref#1{\href{http://arxiv.org/abs/#1}{#1}} % in bibliography
\usepackage{multicol}
\usepackage{color}
\definecolor{rosso}{cmyk}{0,1,1,0.4}
\definecolor{rossos}{cmyk}{0,1,1,0.55}
\definecolor{rossoc}{cmyk}{0,1,1,0.2}
\definecolor{blu}{cmyk}{1,1,0,0.3}
\definecolor{blus}{cmyk}{1,1,0,0.6}
\definecolor{bluc}{cmyk}{1,1,0,0.1}
\definecolor{verde}{cmyk}{0.92,0,0.59,0.25}
\definecolor{verdec}{cmyk}{0.92,0,0.59,0.15}
\definecolor{verdes}{cmyk}{0.92,0,0.59,0.4}

\font\tenrsfs=rsfs10 at 12pt
\font\sevenrsfs=rsfs7
\font\fiversfs=rsfs5
\newfam\rsfsfam
\textfont\rsfsfam=\tenrsfs
\scriptfont\rsfsfam=\sevenrsfs
\scriptscriptfont\rsfsfam=\fiversfs
\def\mathscr#1{{\fam\rsfsfam\relax#1}}

\def\baselinestretch{0.97}

\def\circa#1{\,\raise.3ex\hbox{$#1$\kern-.75em\lower1ex\hbox{$\sim$}}\,}

\newcommand{\beq}{\begin{equation}}
\newcommand{\eeq}{\end{equation}}

\def\circa#1{\,\raise.3ex\hbox{$#1$\kern-.75em\lower1ex\hbox{$\sim$}}\,}
\makeatletter

\def\art{\@ifnextchar[{\eart}{\oart}}
\def\eart[#1]#2#3#4#5#6{{\rm #2}, {#3 #4} {\rm (#6) #5} [{\hhref{#1}}]}
\def\hepart[#1]#2{{\rm #2, \hhref{#1}}}
\newcommand{\oart}[5]{{\rm #1}, {#2 #3} {\rm (#5) #4}}

\newcounter{alphaequation}[equation]
\def\thealphaequation{\theequation\hbox to
0.6em{\hfil\alph{alphaequation}\hfil}}
\def\eqnsystem#1{
\def\@eqnnum{{\rm (\thealphaequation)}}
\def\@@eqncr{\let\@tempa\relax \ifcase\@eqcnt \def\@tempa{& & &} \or
  \def\@tempa{& &}\or \def\@tempa{&}\fi\@tempa
  \if@eqnsw\@eqnnum\refstepcounter{alphaequation}\fi
\global\@eqnswtrue\global\@eqcnt=0\cr}
\refstepcounter{equation} \let\@currentlabel\theequation \def\@tempb{#1}
\ifx\@tempb\empty\else\label{#1}\fi
\refstepcounter{alphaequation}
\let\@currentlabel\thealphaequation
\global\@eqnswtrue\global\@eqcnt=0 \tabskip\@centering\let\\=\@eqncr
$$\halign to \displaywidth\bgroup \@eqnsel\hskip\@centering
$\displaystyle\tabskip\z@{##}$&\global\@eqcnt\@ne
\hskip2\arraycolsep\hfil${##}$\hfil& \global\@eqcnt\tw@\hskip2\arraycolsep
$\displaystyle\tabskip\z@{##}$\hfil
\tabskip\@centering&\llap{##}\tabskip\z@\cr}
\def\endeqnsystem{\@@eqncr\egroup$$\global\@ignoretrue} \makeatother

\def\FERMI{{\sf Fermi}}
\def\PAMELA{{\sf PAMELA}}
\def\HESS{{\sf H.E.S.S.}}
\def\MAGIC{{\sf MAGIC}}

\long\def\symbolfootnote[#1]#2{\begingroup\def\thefootnote{\fnsymbol{footnote}}
\footnote[#1]{#2}\endgroup}

\definecolor{rossoCP3}{cmyk}{0,.88,.77,.40}
\definecolor{verdeCP3}{rgb}{0.09765625, 0.57421875, 0.1015625}
\definecolor{bluCP3}{rgb}{0, 0.23, 0.67}

\begin{document}

\begin{titlepage}
 \begin{center}
{{\LARGE {\color{rossoCP3}
\bf    Gamma Ray Constraints  \\~ on ~\\Flavor Violating Asymmetric Dark Matter
 \rule{0pt}{25pt}}
}} 
 \end{center}
 \par \vskip .2in \noindent
\begin{center}
{\sc {\color{black}Isabella Masina$^{a,b}$\!\!\symbolfootnote[1]{masina@fe.infn.it}, Paolo Panci$^a$\!\!\symbolfootnote[3]{panci@cp3-origins.net},  Francesco Sannino$^a$\!\!\symbolfootnote[4]{sannino@cp3.dias.sdu.dk}}
}
\end{center}
\begin{center}
  \par \vskip .1in \noindent
\mbox{\it
$^a$ CP$\,^3$-Origins \& DIAS,
University of Southern Denmark,
Odense, Denmark}
   \par \vskip .1in \noindent
   \mbox{\it
$^b$   Dip.~di Fisica dell'Universit\`a di Ferrara \& INFN Sez.~di Ferrara,  Ferrara, Italy}
   \par \vskip .4in \noindent
\end{center}
\begin{center}{\large Abstract}\end{center}
\begin{quote}
We show how cosmic gamma rays can be used to constrain models of asymmetric Dark Matter decaying into lepton pairs by violating flavor. First of all we require the models to explain the anomalies in the charged cosmic rays measured by \PAMELA, \FERMI\ and \HESS; performing combined fits we determine the allowed values of the Dark Matter mass and lifetime.  For these models, we then determine the constraints coming from the measurement of the isotropic $\gamma$-ray background by \FERMI\ for a complete set of lepton flavor violating primary modes and over a range of DM masses from 100 GeV to 10 TeV. We find that the \FERMI\ constraints rule out the flavor violating asymmetric Dark Matter interpretation of the charged cosmic ray anomalies. 
\\
[.33cm]
{%\footnotesize
\small \it {Preprint: CP$\,^3$-Origins-2012-014 \& DIAS-2012-15}}
 \end{quote}
 
\vfill

 \end{titlepage}

 %        \newpage
%\def\baselinestretch{1.6}

\def\baselinestretch{1.0}
\tiny
\setlength{\unitlength}{1mm}
%\begin{fmffile}{diagrammi}

%\noindent{\bf \vvv{green} = new}
%\\
%{\bf \rrr{red} = corrections}
%\\
%{\bf \colorbox{yellow}{highlighted} = to be implemented}

\normalsize

%\tableofcontents

\section{Introduction}
\label{Intro}

It is interesting to explore as model independent as possible all the relevant properties of Dark Matter (DM). In practice, due to the different experimental setups, we can only explore a limited set of DM properties. For example, for indirect DM searches, till recently one could explore whether DM is decaying or purely annihilating; and on the basis of its standard model couplings, hope to determine its mass and decay rate and/or annihilation cross section. 

Another interesting general property of DM deals with the possibility for it to violate flavor. The consequences of flavor violating DM in the quark sector have been investigated in~\cite{Kile:2011mn,Kamenik:2011nb,Agrawal:2011ze} considering specific signatures at colliders. In this paper we study the impact of such models in indirect DM searches focussing on the lepton sector which is less constrained. 
%Another interesting general property of DM deals with the possibility for it to violate flavor. 
In particular, we have recently shown that the charged cosmic ray experiments are not yet enough sensitive to measure a possible charge asymmetry in the electron and positron fluxes \cite{Frandsen:2010mr,Masina:2011hu,Chang:2011xn}. If the ratio of the electron to positron fluxes will be found to differ from unity and this were to be attributed to a primary decaying DM component, one would learn that such DM component should necessarily be asymmetric and lepton flavor violating \cite{Masina:2011hu}. We will refer to this possibility as A(symmetric) F(lavor) V(iolating) dark matter (AFVdm).   
%Interestingly it is sufficient to use the information contained in the experimental positron fraction as well as the 
%independent electron and positron fluxes combined with the total $(e^{+} + e^{-})$ flux to start investigating this new general DM  property. 
%Of course, the total flux is flavor blind and therefore the AFVdm information is only contained in the other observables. 
%This warrants a more detailed phenomenological study of AFVdm using the current data. 
The aim of this work is to consider the impact of gamma ray measurements on AFVdm models.

\smallskip
It is useful therefore to summarize the experimental landscape which sets the scene for our phenomenological investigation. The data  collected by \PAMELA \cite{PAMELApositrons1, PAMELApositrons2} and recently by \FERMI \cite{FERMIpositrons}
indicate that there is a positron excess with a rising behavior in the cosmic ray (CR) energy spectrum above $10$ GeV. On the other hand \PAMELA's data  show no unexpected features in the protons nor the anti-protons fluxes \cite{Adriani:2010rc, Adriani:2011cu}. \FERMI\ \cite{Abdo:2009zk} and \HESS \cite{HESSleptons, HESSleptons2} reported a slight additional harder component, in the total $(e^{+} + e^{-})$ spectrum, on the top of a smooth astrophysical spectrum with eventually a steepening at energies of a few TeV. \FERMI\ has also recently presented updated measurements of the total flux of $(e^{+} + e^{-})$ \cite{Ackermann:2010ij} and of the separate $e^+$ and $e^-$ contributions \cite{FERMIpositrons} confirming the notorious rise exposed by \PAMELA\   in 2008. Recent data of the \PAMELA\ collaboration \cite{Adriani:2011xv} on the $e^-$ flux and of the \MAGIC\  collaboration \cite{Tridon:2011dk} on the total $(e^{+} + e^{-})$ flux
have also been released.

These interesting features in charged CR have drawn much attention, and many explanations have been proposed: For example, these excesses could be due to an inadequate account of the cosmic ray astrophysical background in previous modeling; They could be due to the presence of new astrophysical sources; They could also originate from annihilations and/or decays of leptophilic dark matter particle. The interpretation in terms of dark matter annihilations often leads to an unobserved excess of  neutral messenger probes (essentially gamma rays, but also neutrinos) originating from dense DM concentrations. The interpretation instead in terms of dark matter decays~\cite{NSS, CT, YYLZ, IMM, IT, CNTY, ADDG, ITW} is less constrained, coming from the fact that in this case, the signal is linear in the DM number density. On the other hand, both \FERMI\ \cite{FERMIexg} and \HESS\ telescopes are making huge progress in the study of the gamma ray map constructed by observing more targets in different regions of the sky, including those of interest for decaying DM.

\smallskip
In this work, we first reconsider the interpretation of the charged CR data anomalies as due to AFVdm decaying in lepton pairs, finding
the allowed values for the DM mass and lifetime. 
Then we investigate the impact that the recent \FERMI\ isotropic gamma ray flux  measurement \cite{FERMIexg} has on constraining our
AFVdm models.  It is known that this measurement is a powerful probe for any model involving decaying 
DM (see e.g.~\cite{Takayama:2000uz, Overduin:2004sz, Bertone:2007aw}). 
In our analysis, we use the most advanced semi-analytic tools for DM indirect searches which include the electroweak corrections 
for the primary fluxes and a refinement of the propagation scheme for the $e^{\pm}$. 
As happens for models respecting flavour symmetry, we find that also models of AFVdm decaying into lepton pairs  are not compatible with photon observations.

\section{Flavor Violating  Dark Matter Obervables}

%Therefore, only half of the decays allowed at the Lagrangian level are dominant, i.e. the ones deriving from the surviving component. 

%In this investigation we assume dark matter to couple only to leptons. 

%The scalar interactions we consider are therefore: 
%\begin{equation}
%c_{\ell \ell'}\,T \overline{\ell} \ell' + {\rm h.c.} \ .
%\label{SDM}
%\end{equation}
%Here  $\ell= e, \mu$ or $\tau$ and we assume summation over the standard model leptonic flavor 
%indices. A generic $c_{\ell \ell'}$ leads to violations of the lepton numbers. We assume that, 
%during the evolution of the universe, an asymmetry in the relic densities of $T$ and $T^*$ arises. 
%We further consider the case in which $T^*$ has disappeared and that we are left today only with $T$. 
%The latter decays via the first interaction term given in \eqref{SDM}.
%Explicitly, this leads to 
%\beq
%T \rightarrow \ell^-_{L} {\ell'}^+_{L}  + \ell^-_{R} {\ell'}^+_{R} ~~,
%\label{TLL}
%\eeq
%where the notation means that leptons in the pair are produced with the same helicity 
%and that there is equal probability for both chiralities. Parity is thus conserved.
%If $\ell \neq \ell'$ then asymmetric dark matter implies charge-conjugation violation in the decay.
%Clearly, if $\ell$ ($\ell'$) is not directly the flavor $e$, electrons (positrons) are produced
%in its decay chain. If we were to have symmetric type dark matter then, as it is clear from \eqref{TLL}, 
%we would have an equal energy spectrum of electron and positrons since they would be produced via 
%$T$ and $T^{*}$. 

It is useful to define the flux of charged particles at the Earth location,   coming from a general primary decay mode DM$\rightarrow ij$, as follows: 
\begin{eqnarray}
\Phi^{\rm tot}_{ij}(E)& = &  {\Phi^{e^+}(E)+\Phi^{e^{-}}(E)}  = 
 {\Phi^{e^+}_{i}(E)+\Phi^{e^{-}}_{j}(E)} +  \Phi^{e_s}_{ij}(E) \ , \\
  \Phi^{e_s}_{ij}(E) & = &
 {\Phi^{e_s^+}_{i}(E)+\Phi^{e_s^{-}}_{j}(E)} +  {\Phi^{e_s^+}_{j}(E)+\Phi^{e_s^{-}}_{i}(E)}  \ , 
\end{eqnarray}
where $i$ and $j$ refer to primary charged lepton ($e, \mu, \tau$) with the first index positively charged and the second one negatively charged. $e_s$ stands for the soft part of the spectrum which arises from secondary hadronic decays as can be deduced by comparing PYTHIA against HERWIG for the version not including $\ell \rightarrow \ell \gamma$ and $\gamma \rightarrow f \bar{f}$ with $f$ generic SM fermions and $\ell$ are the charged leptons. In the bottom-left panel of Fig.~2 in \cite{PPPC4DMID} one can see the emergence of the secondary electron and positrons when comparing the two MonteCarlos for the same process. 
We also note that within the standard model the following is an excellent approximation: 
\begin{equation}
 \Phi^{e_s}_{ij}(E)  = 
   2 \Phi^{e_s}_{i}(E)+ 2\Phi^{e_s}_{j}(E)  \ , 
\end{equation}
with $\Phi^{e_s^{+}}_{i}(E) = \Phi^{e_s^{-}}_{i}(E) \equiv \Phi^{e_s}_{i}(E)$ for any $i$. 

  In the case where the AFVdm model violates lepton flavor maximally $i$ is always different from $j$. Following  \cite{Masina:2011hu}  it is  convenient to define the following ratio:  
\begin{equation}
r_{ij}(E) = \frac{\Phi^{e^{-}}_{j}(E)+\Phi^{e_s^{-}}_{j}(E)+\Phi^{e_s^{-}}_{i}(E)}
{\Phi^{e^+}_{i}(E) +\Phi^{e_s^{+}}_{i}(E)+ \Phi^{e_s^{+}}_{j}(E)} =  \frac{2\Phi^{e^{-}}_{j}(E)+\Phi^{e_s}_{ij}(E)}
{2\Phi^{e^+}_{i}(E) +\Phi^{e_s}_{ij}(E)} \ .
\label{ratio}
\end{equation}
We are assuming for the analysis a given $i$ and $j$ to be the dominant decay mode. It is, however, straightforward to generalize the result to include different branching ratios in different channels. 

The case of lepton-flavor preserving primary channels (i.e. $i=j$) has been studied in \cite{CPS1, CPS2, Papucci:2009gd, Zaharijas:2010ca, Zimmer:2011vy, Ke:2011xw, Luo:2011bn, Huang:2011xr}.  An improved analysis appeared recently in \cite{CPS2} making use of the electroweak corrections for the primary fluxes of the relevant stable standard model particles determined in \cite{PPPC4DMID, EWcorr2} and a refinement of the propagation scheme for the $e^{\pm}$  (see \cite{PPPC4DMID} for further details).  In  \cite{EWcorr1} it was already pointed out that the electroweak corrections were relevant. 
   In the case of flavor preserving DM there is the obvious simplification that $\Phi^{e^+}_i=\Phi^{e^-}_i  $ and therefore $\Phi^{\rm tot}_{ii}(E) = 2{\Phi^{e^+}_{i}(E)} +   \Phi^{e_s}_{ii}(E)$. 

In our case we should determine $\Phi^{e^+}_i$ and $\Phi^{e^-}_j$ independently since the  observables useful to identify AFVdm  are sensitive to these independent fluxes. In general it is not possible to use directly  the tools provided in \cite{PPPC4DMID} for AFVdm since in the latter the direct and secondary contributions are not separated. However, as we shall explain, for the decay modes relevant here we can still capitalize on Ref. \cite{PPPC4DMID} while for a more general flavor violation involving also the quarks this is not obvious and should be implemented in the numerical codes. For the total flux of positron and electrons coming from a two-body decays one can always use  the identity: 
\begin{equation}
\Phi^{\rm tot}_{ij}(E)  = \frac{\Phi^{\rm tot}_{ii}(E)+\Phi^{\rm tot}_{jj}(E)}{2} \ .
\end{equation}
%{\color{blue} with the right-hand side fluxes determined using the propagating fluxes from reference \cite{Cirelli:2010xx} $\rightarrow$ lo diciamo sotto, quindi leveri questa frase!!!!!}. 
For the ratio $r_{ij}(E)$ the situation instead is trickier and we make the following phenomenologically excellent approximation valid for scalar DM decaying into two-body lepton-violating processes:
\begin{equation}
r_{ij}(E) = 
\frac{\Phi^{\rm tot}_{jj}(E) +2(\Phi^{e_s^{-}}_{i}(E) - \Phi^{e_s^{-}}_{j}(E))
}{{\Phi^{\rm tot}_{ii}(E)} + 2(\Phi^{e_s^{+}}_{j}(E) - \Phi^{e_s^{+}}_{i}(E))} \approx  \frac{\Phi^{\rm tot}_{jj}(E)  
}{{\Phi^{\rm tot}_{ii}(E)}} \ ,
\end{equation}
where we neglected the differences which are numerically very small. This quantity can be directly related to the measured positron fraction (see~\cite{Masina:2011hu}) representing the relevant observable for AFVdm. 
 
%
%{\color{blue} In this paper, we will focus on a complex scalar DM candidate,  assuming it to be of asymmetric type. Similar results and considerations can be applied for other cases studied in~\cite{Frandsen:2010mr, Masina:2011hu} (i.e. vector DM candidate), since the current accuracy of the positron fraction datapoints does not allowed a discrimination among them. }
% \section{Fits to CR anomalies and $\gamma$ ray Constraints}

 \subsection{DM interpretations of the charged CR anomalies}
 \label{DMinterp}

\begin{figure} 
\begin{minipage}{.48\textwidth}
\centering
\includegraphics[width=\textwidth]{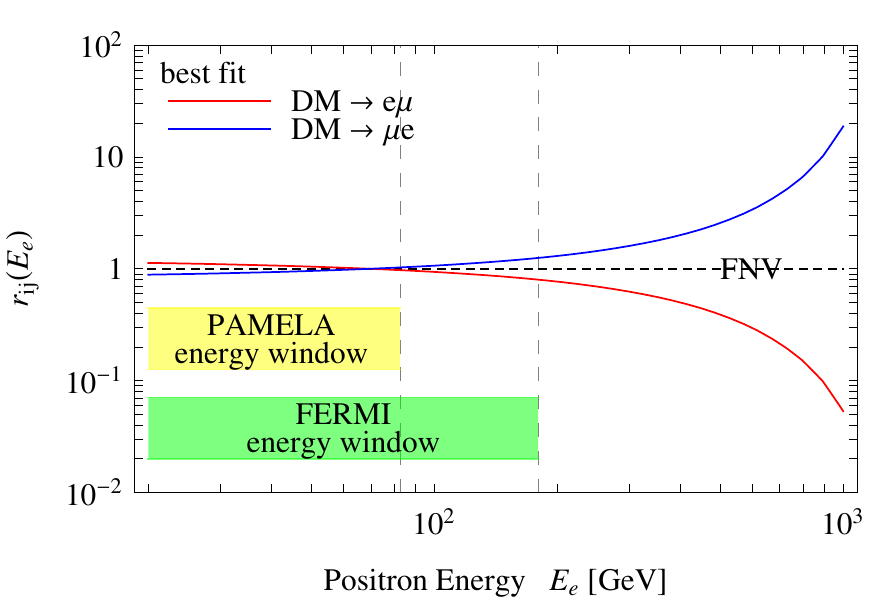} \\
\includegraphics[width=\textwidth]{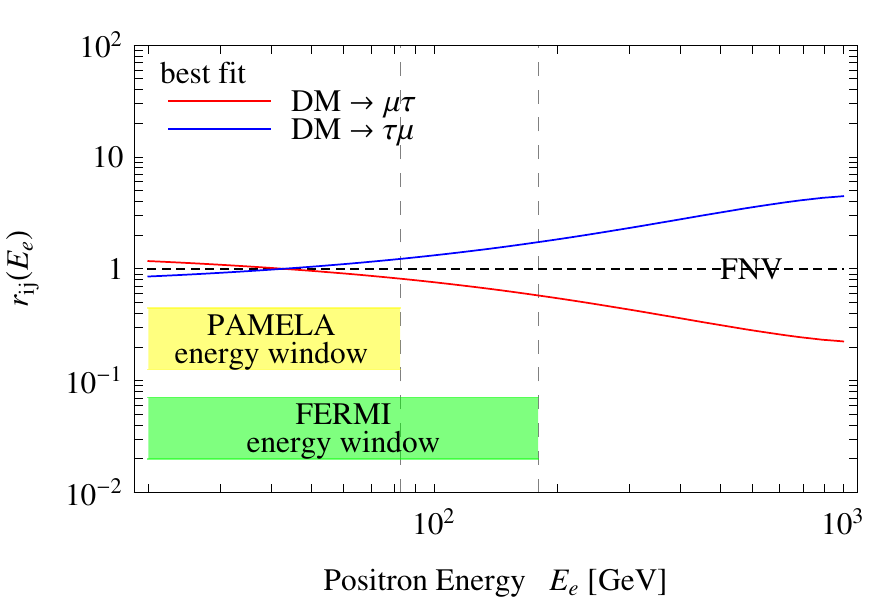}
\end{minipage}
\begin{minipage}{.02\textwidth}
\tiny{.}
\end{minipage}
\begin{minipage}{.48\textwidth}
\vspace{-.4cm}
\centering
\includegraphics[width=\textwidth]{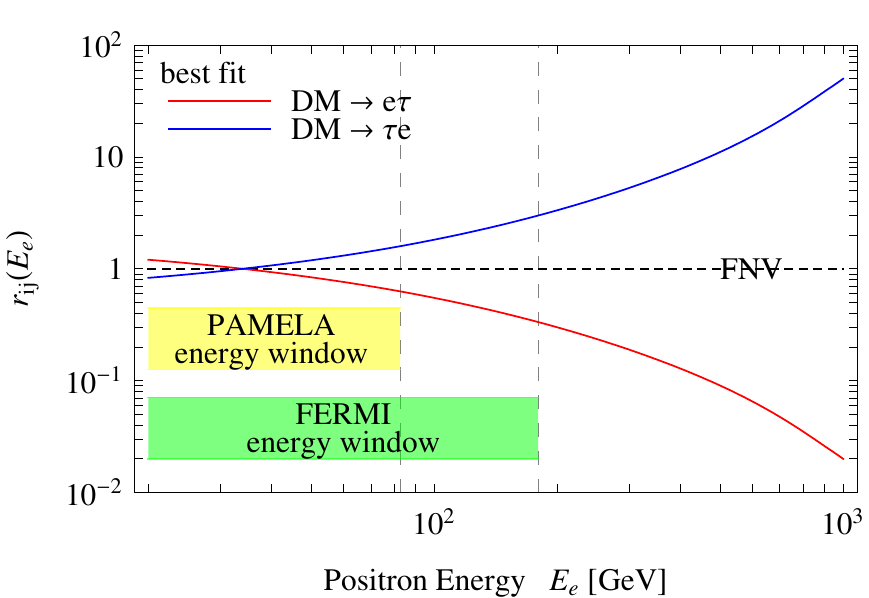} \\
%\vspace{-.1cm}
 \caption{ The ratio $r_{ij}(E)$ as a function of the positron energy. In each plots the blue/red lines are performed  by using the best fit DM mass for the primary channels DM$\rightarrow ij$/DM$\rightarrow ji$. The yellow and green bands represent the energy ranges of the positron fraction data of \PAMELA\ and \FERMI\ respectively. The horizontal dashed lines refer to flavor non-violating models for which $r_{ij}(E)\equiv1$.\label{fig:A}}

\end{minipage}

\end{figure}
 
As already mentioned in Sec.~\ref{Intro}, the anomalous \PAMELA, \FERMI\ and \HESS\ data have been already interpreted in terms of decaying DM models. Here we present a careful analysis of the charged CR anomalies in the context of lepton flavor-violating asymmetric DM. 

\smallskip
 We use the following set of data: The \PAMELA~\cite{PAMELApositrons1,PAMELApositrons2} and \FERMI~\cite{FERMIpositrons} positron fraction  selecting points with  $E> 20$ GeV; The total fluxes ($e^++e^-$) of \FERMI~\cite{Ackermann:2010ij}, \HESS~\cite{HESSleptons, HESSleptons2} and \MAGIC~\cite{Tridon:2011dk}; Finally the \PAMELA\ $\bar{p}$ flux \cite{Adriani:2010rc}.  We select the data from 20 GeV and above for \PAMELA\ in order to:  $i)$  have a more consistent overlap with the \FERMI\ positron fraction data points, $ii)$  to avoid the low-energy region affected by the uncertainty coming from the solar modulation. The total fluxes and positron fraction used to compare with data have been determined using the tools in \cite{PPPC4DMID}, which correspond to the flavor-preserving fluxes $\Phi^{\rm tot}_{ii}(E)$. To determine the best fit parameters for the DM mass ($M_{\rm DM}$) and flavor-violating decay half-life ($\tau_{\rm dec}$) we also varied within the allowed uncertainties: The slope and the normalization of the energy dependent parametrization of the astrophysical background;  The propagation parameter of charged cosmic rays.  We  follow the analysis for the flavor non violating processes  presented in  \cite{CKRS} and \cite{CPS1} where the reader can find more details.  We have checked that the allowed regions of the parameters are not significantly modified even if we consider the recent determination of the $e^+$ astrophysical background given in~\cite{Lavalle:2010sf} rather than the standard one \cite{Moskalenko:1997gh,Baltz:1998xv}). 
 
\smallskip
We have found  that it is possible to obtain a reasonable combined fit of the whole datasets of charged CR for all the leptonic flavor-violating primary modes (DM$\rightarrow ij$) as well as their charge conjugate (DM$\rightarrow ji$). 
This can be appreciated in Fig.~\ref{fig:C}, where we show the allowed regions in the plane ($M_{\rm DM}, \,\tau_{\rm dec}$) for all the channels we consider. In particular the red/blue and orange/magenta blobs refer to the 95.45\% and 99.999\% C.L. regions for the channels DM$\rightarrow ij$/DM$\rightarrow ji$ respectively.
We have also checked that the resulting flux of antiprotons remains within the experimental bounds even  when taking into account the final state radiation of electroweak gauge bosons.   The quality of the fit to the CR data for flavor-violating processes from asymmetric DM is as good as the ones for flavor-preserving processes.  

However, only for flavor-violating process coming from the asymmetric DM sector the ratio $r_{ij}(E)$ can be, and will be, different from unity \cite{Masina:2011hu}. Therefore it is important to show the resulting ratio using the best fit value of the DM mass recalling that by construction $r_{ij}(E)$  does not  depend on the decay rate. The resulting $r_{ij}(E)$ for the different processes  are shown in Fig.~\ref{fig:A}. A ratio larger than unity is associated to an electron flux at Earth bigger, at a given energy, than the positron one. Within the energy ranges investigated by \PAMELA\ (yellow band) and \FERMI\ (green band) for the positron fraction data, whose maximum extent has been marked in the figures by two vertical lines, one can appreciate a sizable deviation from unity only for the $e\tau$ primary mode and its charged conjugate. This last result is in agreement with the preliminary analysis performed in \cite{Masina:2011hu}.  An interesting result from the investigation of the combined allowed region of the parameter space is that for sufficiently large values of $r_{ij}(E)$ one can resolve the degeneracy between $ij$ and $ji$. As one can see from Fig.~\ref{fig:C}, with the current precision of the positron excess data points, the discrimination among red/orange and blue/magenta regions turns out to be experimentally appreciable only for the  $e\tau$ primary mode.
 
As a final comment, it is perhaps worth stressing that in order to be coherent with the \FERMI\ positron energy range data, and to be absolutely sure that we were not affected by the solar modulation we only considered  data starting from 20 GeV. If we were to add also the lower energy data of \PAMELA\  (between 10 and 20 GeV) the two blobs would be even more separated.

 \subsection{Isotropic Gamma Ray Constraints}
 \label{DMgamma}

 \begin{figure}[t] 
\begin{minipage}{.48\textwidth}
\centering
\includegraphics[width=\textwidth]{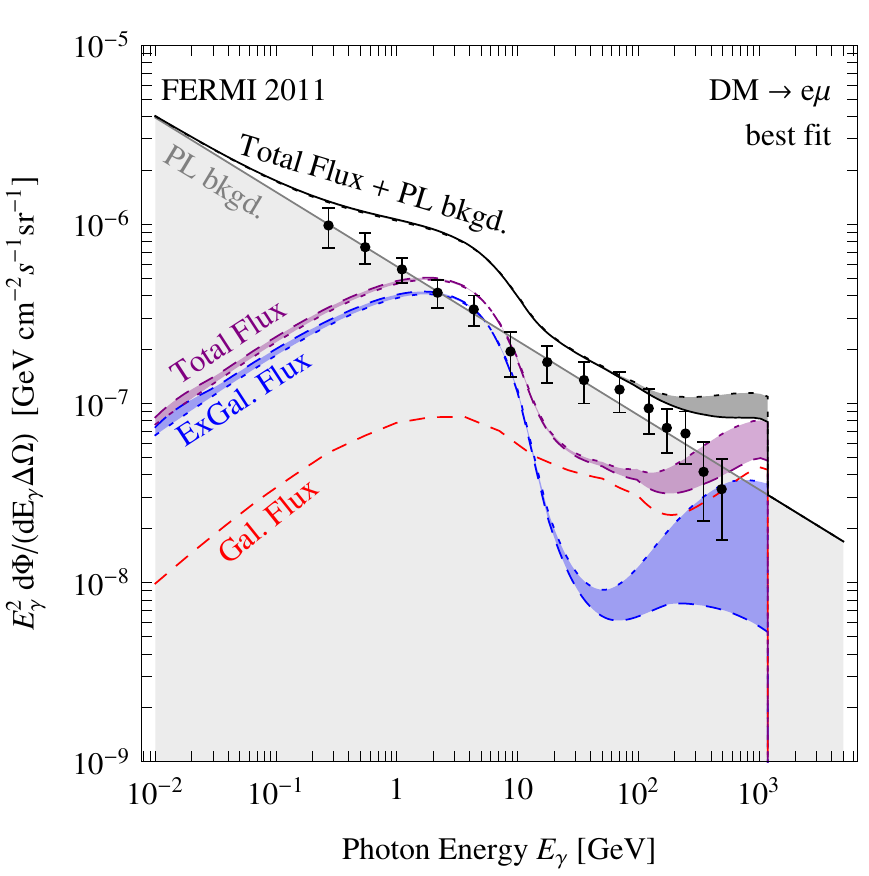} \\
\includegraphics[width=\textwidth]{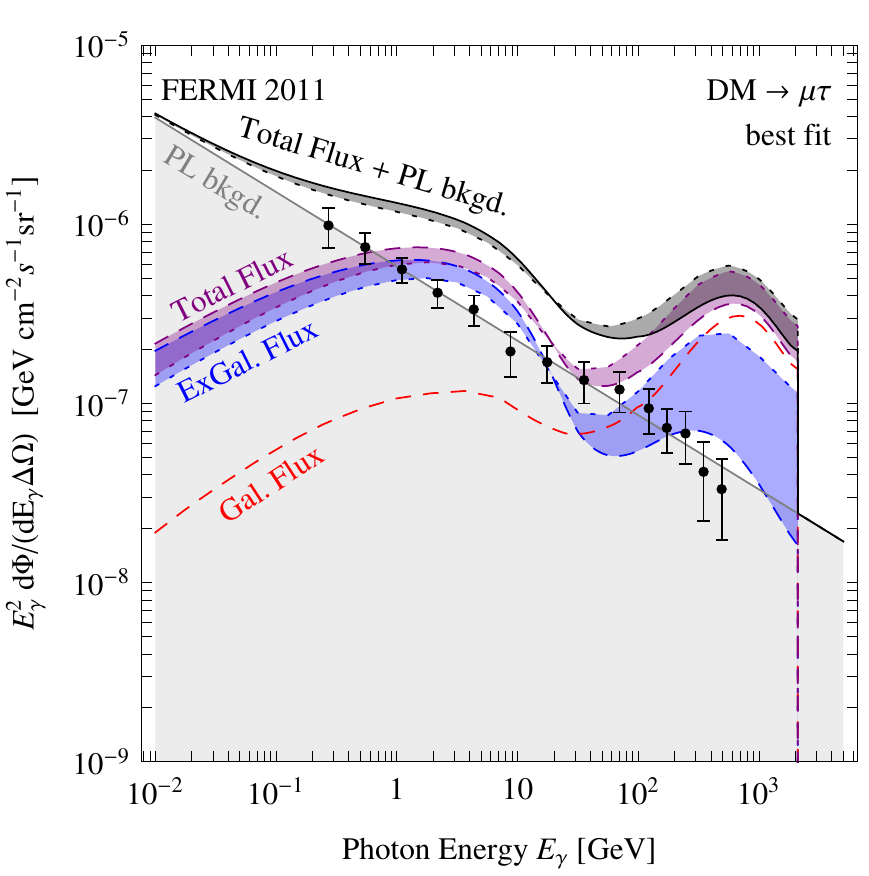}
\end{minipage}
\begin{minipage}{.02\textwidth}
\tiny{.}
\end{minipage}
\begin{minipage}{.48\textwidth}
%\vspace{-.1cm}
\centering
\includegraphics[width=\textwidth]{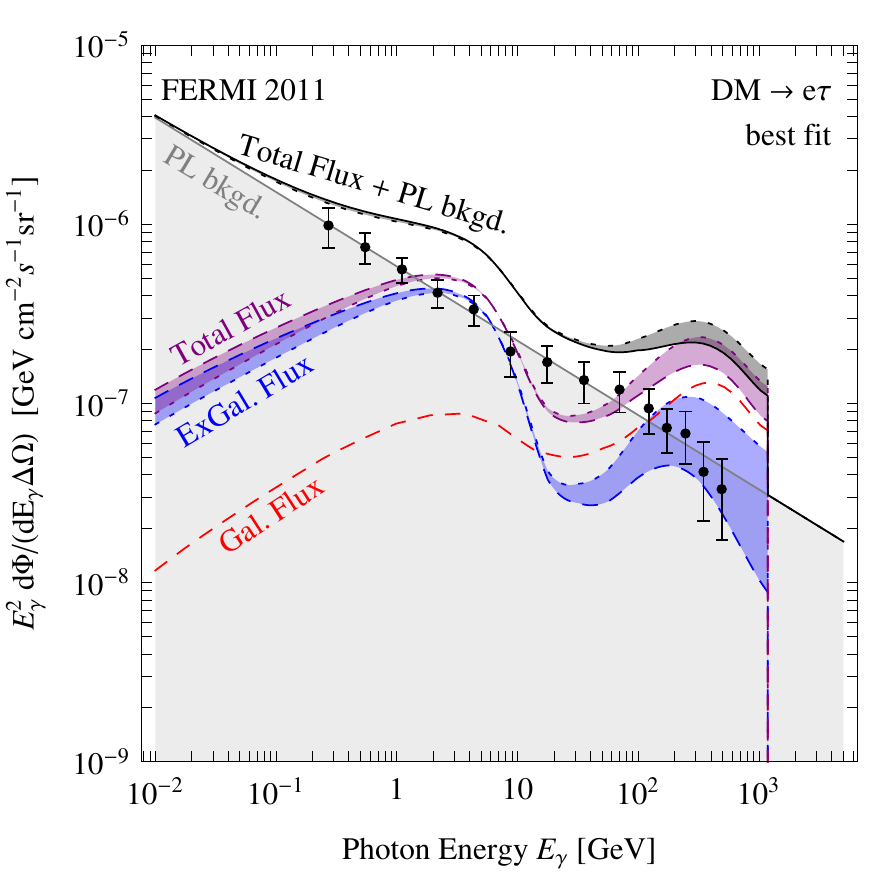} \\
\vspace{-.1cm}
 \caption{  Isotropic $\gamma$-ray signals from an asymmetric flavor-violating decaying DM candidate that fits the charged CR anomalies. The DM contributions are dashed and the astrophysical power law background is shaded gray. In each panel, the dashed red line refers to the ``isotropized'' galactic flux and the dashed/dotted blue lines refer to the extragalactic fluxes with/without the absorption of high energetic gamma rays and redistribution of them at lower energy. The black solid line represents the total DM fluxes with UV absorption plus a power law background that we use for deriving the $\gamma$-ray constraints.\label{fig:B}}

\end{minipage}

\end{figure}

The isotropic residual gamma ray flux measured by \FERMI~\cite{FERMIexg} extends from about $200$ MeV up to $580$ GeV.  This flux is due to a variety of physical phenomena such as unresolved sources as well as genuine diffuse processes (see~\cite{FERMIexg}). 

If DM is present in the sky and it decays it will, in general, also contribute to this isotropic flux:
\begin{equation}  
\frac{d \Phi_{\rm isotropic}}{d E_\gamma} = \frac{d \Phi_{\rm ExGal}}{d E_\gamma} + 
4 \pi \left. \frac{d \Phi_{\rm Gal}}{d E_\gamma\, d \Omega}\right|_{\rm minimum}
\label{eq:main}
\end{equation} 
where the former term is the extragalactic cosmological flux and it is truly isotropic. The latter, which describes the residual emission from the DM halo of our Galaxy it is not, but its minimum constitutes an irreducible contribution to the isotropic flux. Typically the two contributions are comparable.
% extragalactic cosmological origin and is truly isotropic while the second contribution from our own galaxy and its minimum constitutes and irreducible contribution to the isotropic flux. Typically the two contributions are comparable. 
We followed the approach established in  \cite{CPS1, CPS2} where the reader can find further details.

\medskip
Here we need to generalize the situation to the case of $ij$ primary modes for $i$ different from $j$ and write: 

\begin{equation}
\frac{d \Phi^{ij}_{\rm ExGal}}{d E_\gamma}
 = \Gamma_{\rm dec}\, \frac{\Omega_{\rm DM}\,\rho_{c,0}}{M_{\rm DM}} 
\,\int_0^\infty d z\,\frac{e^{-\tau(E_\gamma(z),z)}}{H(z)} \frac12  \left[\frac{d N^{ii}}{d E_\gamma}(E_\gamma(z),z) + \frac{d N^{jj}}{d E_\gamma}(E_\gamma(z),z) \right]\,,
\label{smoothedmap}
\end{equation}
where $\frac{d N^{ii}}{d E_\gamma}(E_\gamma(z),z) $ are taken from the non flavor violating processes and photon energy is $E_\gamma$. The formula above makes use of the fact that the gamma flux for the case studied here corresponds to a DM decaying without violating flavor into 50\% $ii$ and $jj$ modes.  It is therefore clear that the photon signal as well as the total flux of positron and electron alone cannot discriminate between AFVdm  and flavor symmetric DM.
Here  $\Gamma_{\rm dec}= (\tau_{\rm dec})^{-1}$ is the decay rate and  $H(z) = H_0\sqrt{\Omega_M(1+z)^3+\Omega_\Lambda}$ is the Hubble function where $H_0$ is the present Hubble expansion rate. The DM, matter and cosmological constant energy density are $\Omega_{\rm DM}$, $\Omega_M$  and $\Omega_\Lambda$  and are expressed in units of the critical density, $\rho_{c,0}$.  At any redshift $z$, the  spectrum of gamma ray $d N/d E_\gamma$ is given by: a) the prompt $\gamma$-ray emission from DM decays and b) the Inverse Compton Scatterings (ICS) on CMB photons of the $e^+$ and $e^-$ from the same decays. Using now Eq.~(\ref{smoothedmap}), we can compute the extragalactic flux in terms of known quantities for any specified DM mass $M_{\rm DM}$ and flavor-violating decay  channel. The resulting fluxes are taken from~\cite{PPPC4DMID} with an improved  treatment of the optical  depth (the factor $e^{-\tau(E_\gamma(z),z)}$ in Eq.~(\ref{smoothedmap})) in which the absorption of high energy gamma rays due to scattering with the extragalactic UV background light is fully taken into account. As recently pointed out in~\cite{CPS2}, this effect is important at energies $E_\gamma \gtrsim 100\,$GeV and can decrease the flux at high energy by about one order of magnitude. In Fig.~\ref{fig:B} we show the fluxes obtained neglecting absorption (dotted blue lines) and the one including it (dashed blue lines). As one can see the effect is sizable for all the channels we consider. In particular for the processes which involved a $\tau$ in the primary modes one can even appreciate the redistribution of high energy $\gamma$-ray towards the lower part of the spectrum. This will affect the $\gamma$-ray constraints especially for channels which feature a large prompt contribution. In deriving the $\gamma$-ray constraints we will take into account the full UV absorption.

\smallskip
In our case, the galactic differential flux coming from a given direction of the sky $d \Omega$ is:  
\begin{equation}
\frac{d \Phi^{ij}_{\rm Gal}}{d E_\gamma \ d \Omega} = \frac{1}{4\pi} \frac{\Gamma_{\rm dec}}{M_{\rm DM}} \int_{\rm{los}} d s \, \rho_{\rm halo}[r(s,\psi)] \,\frac12  \left[\frac{d N^{ii}}{d E_\gamma}  + \frac{d N^{jj}}{d E_\gamma}  \right]\,,
\label{fluxdec}
\end{equation}
where the coordinate $s$ parameterizes the distance from the Sun along the line-of-sight (los).  
Here $\rho_{\rm halo}$ is the Milky Way DM distribution, for which we take the standard Navarro-Frenk-White~\cite{Navarro:1995iw} profile. One can consider in principle other choices of DM profiles, however, for decaying DM these choices lead to comparable results.  The coordinate $r$, centered on the galactic center (GC), reads $r(s,\psi)=(r_\odot^2+s^2-2\,r_\odot\,s\cos\psi)^{1/2}$, where $r_\odot = 8.33$ kpc is the most likely distance of the Sun from the GC and $\psi$ is the angle between the direction of observation in the sky and the GC.  $d N/d E_\gamma$ is again the sum of two components: the prompt one and the ICS one. They  are then determined by using the tools in~\cite{PPPC4DMID}. In particular, for the ICS flux we 
use the full spatial dependence of the energy losses, which can be taken into account by a generalized halo functions for the IC radiative process provided there. In finding the minimum of the galactic flux we make the reasonable approximation that it corresponds to the value at the anti galactic center  \cite{CPS2}.  
In formul\ae
\begin{equation}  
\left. \frac{d \Phi_{\rm Gal}}{d E_\gamma\, d \Omega}\right|_{\rm minimum} \rightarrow \left. \frac{d \Phi_{\rm Gal}}{d E_\gamma\, d \Omega}\right|_{\rm anti-GC} \ .
\label{eq:main}
\end{equation} 

In Fig.~\ref{fig:B} we show the ``isotropized'' galactic flux for all the channels we consider (dashed red lines). For all the panels one observes the low energy contribution  due to  ICS on CMB photons, and the high energy one coming from prompt emission. The junction between the two parts of the spectrum originates from the ICS on the residual starlight and infrared light present along the los towards the anti galactic center.  
   
\begin{figure}[t] 
\begin{minipage}{.48\textwidth}
\centering
\includegraphics[width=\textwidth]{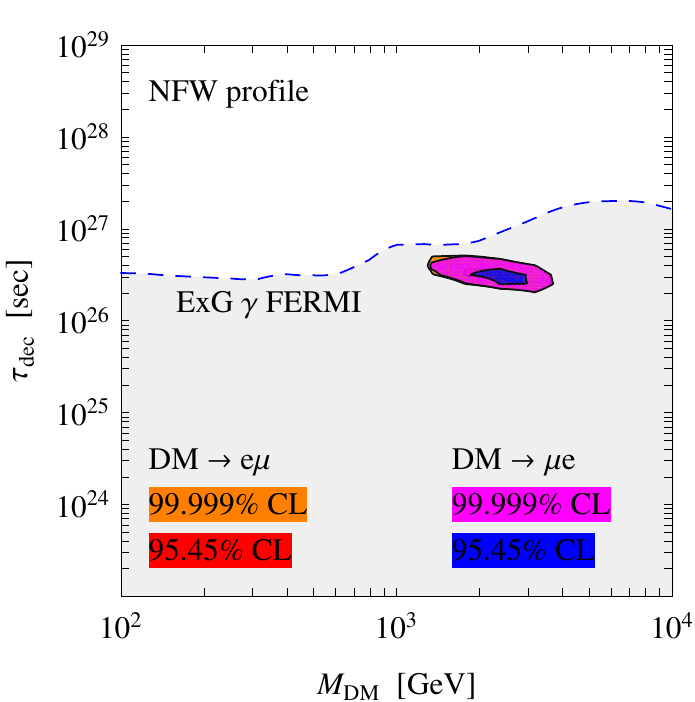} \\
\includegraphics[width=\textwidth]{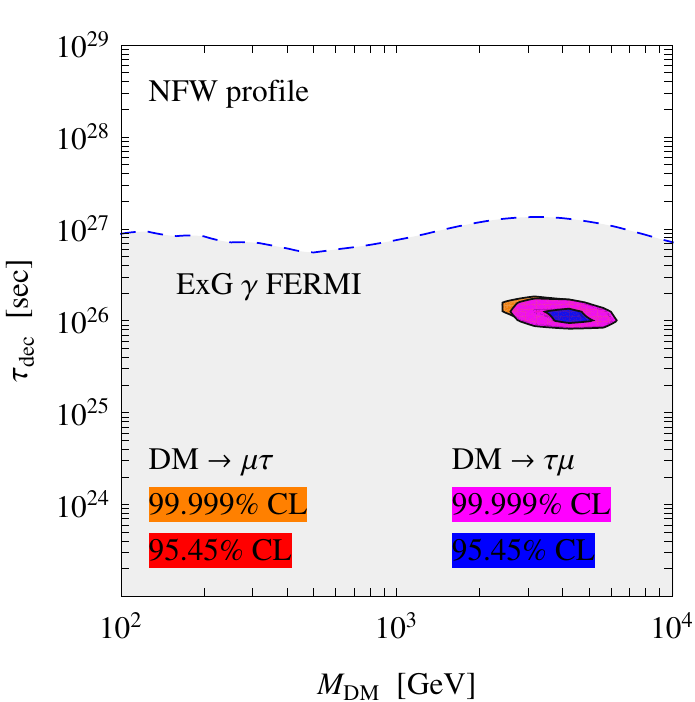}
\end{minipage}
\begin{minipage}{.02\textwidth}
\tiny{.}
\end{minipage}
\begin{minipage}{.48\textwidth}
\vspace{-1.75cm}
\centering
\includegraphics[width=\textwidth]{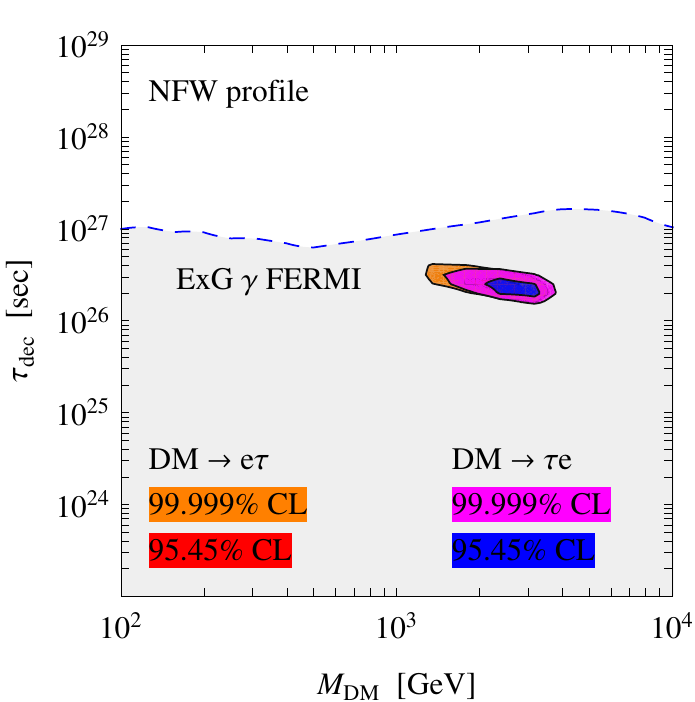} \\
\vspace{.75cm}
 \caption{The regions on the parameter space ($M_{\rm DM},\, \tau_{\rm dec}$) that are excluded by the isotropic $\gamma$-ray measurement  of \FERMI\  (blue dashed lines) together with the regions of the global fit to the charged CR data, for different flavor-violating decay channels. In each panel  the red/blue and orange/magenta blobs refer to the 95.45\% and 99.999\% C.L. allowed regions for the channels DM$\rightarrow ij$/DM$\rightarrow ji$ respectively. \label{fig:C}}

\end{minipage}

\end{figure}   
   
\medskip   
   
We are now able to compare the isotropic gamma ray flux coming from AFVdm with the one measured by \FERMI. In Fig.~\ref{fig:B} we show an example for the candidates which fit the charged CR anomalies. It is clear that the signal overshoots the data and even provides local, in energy, features which are clearly not present in the data. In fact, the data are in good agreement with an astrophysical power-law background. To determine the actual constraints we demand that the sum of the DM signal and power-law  background  (solid black lines in Fig.~\ref{fig:B})  does not exceed a given significance.  There are several potential sources that may produce such background, like unresolved blazars, star-forming galaxies or electromagnetic cascades from ultrahigh-energy cosmic ray losses. In general, a combination of them is not expected to produce an exact power law, which could be also inaccurate for some of the different contributions taken alone. However, within the actual precision of the detector, we find that the data alone (without DM contribution) are well fitted by a power-law index -2.41 and normalization $1.02\times 10^{-5}$/(GeV\,cm$^2$\,sr). In view of that and since the bumpy shape of the DM signal is so different from the featureless observations, it is reasonable to consider a power-law astrophysical background  for the computations of the constraints on DM properties. %Furthermore when we compute the bounds, we let the normalization of the power-law background vary within a factor of 2 from the central value specified above and its index within 0 and -1. This choices are asto-physically plausible, although varying the parameters in a broader range would not change the results shown in our paper.
%A power-law astrophysical background can be justified as emerging when considering different astrophysical sources. We find that the \FERMI\ data are well described by a power law with index -2.41 and normalization 1.02 $\, 10^{-5}$ GeV/(cm$^2$ sec sr). 
On the top of this power-law, for a given $M_{\rm DM}$ we add the DM signal, whose normalization is controlled by $\Gamma_{\rm dec}$, letting the normalization and the index of the DM free power-law vary\footnote{When we compute the bounds, we let the normalization of the power-law background to
vary within a factor of 2 from the central value specified above, and its index within 0 and -1. These choices are astro-physically plausible, although varying the parameters in a broader range would not change the results.}  and marginalizing over them. We then compute the $\chi^2$ and impose 95\% C.L. limits on $\Gamma_{\rm dec}$.

\medskip
 The exclusion plots can be found in Fig.~\ref{fig:C} where we summarize our results for AFVdm. As one can see the isotropic $\gamma$-ray data measured by the \FERMI\ satellite exclude DM lifetimes of the order of $10^{26}$ to few $10^{27}$ seconds. This therefore rules out the interpretation of the charged CR excesses in terms of  AFVdm. In this study we have focussed on two-body lepton flavor-violation decay modes of scalar asymmetric DM.  For vector asymmetric DM, the decay channels  differ only in  polarization with respect to the ones coming from scalar DM. Therefore, since this differences are not large, it is reasonable to expect that the allowed regions and constraints showed in Fig.~\ref{fig:C} apply also for these cases.

%%%%%%%%%%%%%%%%%%%%%%%%%%%%%%%%%%%%%%%%%%%%%%%%%%%%%%%%

\section{Conclusions}

 Flavor violating decaying asymmetric DM models allow for a new type of DM phenomenology and can feature heavy or/and light DM candidates. Here we showed that it is possible to explain the anomalies in the charged CR measured by \PAMELA, \FERMI\ and \HESS\ in the context of AFVdm.  Because of the nature of AFVdm  the ratio of the electron  over positron DM induced flux is, in general, different from unity while it must be unity for any other type of DM. We therefore determined the associated  energy and flavor dependent ratio $r_{ij}(E)$ from the combined fits to the CR anomalies and found that the current data allow for flavor-violation for asymmetric DM  for all the decay processes.  An interesting result from the investigation of the combined allowed region of the parameter space is that for sufficiently large values of $r_{ij}(E)$ one could resolve the degeneracy between $ij$ and $ji$. 

We then discussed the constraints coming from the measurement of the isotropic $\gamma$-ray background by \FERMI\ for a complete set of lepton flavor-violating primary modes and over a range of DM masses from 100 GeV to 10 TeV.  We used updated computational tools which are the most refined semi-analytical computations present in the literature.  

We found that \FERMI\ constraints rule out the AFVdm interpretation of the charged CR anomalies. This analysis complements the constraints for flavor preserving DM for which similar constraints have been derived.

\bigskip

\noindent {\bf Acknowledgments.}   
We thank Marco Cirelli and Alejandro Ibarra for useful discussions.
 
  %%%%%%%%%%%%%%%%%%%%%%%%%%%%%%%%%%%%%%%%%%%%%%%%
%\newpage

\clearpage

\footnotesize{
\begin{multicols}{2}
  
\end{multicols}}

\end{document}